# The Future of Computing: Bits + Neurons + Qubits

Dario Gil and William M. J. Green

IBM Thomas J. Watson Research Center, Yorktown Heights, NY 10598

**Abstract**

The laptops, cell phones, and internet applications commonplace in our daily lives are all rooted in the idea of zeros and ones – in bits.

This foundational element originated from the combination of mathematics and Claude Shannon's Theory of Information. Coupled with the 50-year legacy of Moore's Law, the bit has propelled the digitization of our world.

In recent years, artificial intelligence systems, merging neuron-inspired biology with information, have achieved superhuman accuracy in a range of narrow classification tasks by learning from labelled data. Advancing from Narrow AI to Broad AI will encompass the unification of learning and reasoning through neuro-symbolic systems, resulting in a form of AI which will perform multiple tasks, operate across multiple domains, and learn from small quantities of multi-modal input data.

Finally, the union of physics and information led to the emergence of Quantum Information Theory and the development of the quantum bit - the qubit - forming the basis of quantum computers. We have built the first programmable quantum computers, and although the technology is still in its early days, these systems offer the potential to solve problems which even the most powerful classical computers cannot.

The future of computing will look fundamentally different than it has in the past. It will not be based on more and cheaper bits alone, but rather, it will be built upon bits + neurons + qubits. This future will enable the next generation of intelligent mission-critical systems and accelerate the rate of science-driven discovery.

# 1. Introduction

The field of computing is currently experiencing an extraordinarily challenging inflection, characterized by several key disruptive factors.

The previous 60 years of progress in classical computing has been underpinned by Moore's Law and its observation that roughly every two years, integrated circuits scaled to either boost performance or cut cost by 2x. However, the cadence of new transistor technology nodes has slowed in the past decade, and as such, the exponential growth engine driving classical computing will plateau. This slowdown is forcing a transition to heterogeneous system designs augmented with special-purpose accelerators, to address the need for increased performance within traditional cost and power envelopes. However, such systems are significantly more difficult to program and utilize efficiently.

Advances in raw digital computing power have nevertheless brought us to a point where biologically-inspired models of computation based upon neural nets are now highly integrated into the state of the art. Artificial Intelligence (AI) capabilities have grown enormously in their capacity to interpret and analyze data for value extraction in research, consumer, and enterprise applications. However, AI significantly complicates workload development and execution, since it inserts coupled, computationally demanding elements within a workflow, variously performing learning and inferencing steps against the data within the workflow. The complexity of AI workflows is such that large multidisciplinary teams are currently required to manage development and production. At the same time, this complexity presents an intrinsic opportunity for the application of AI to the management and optimization of entire computational systems.

Concurrently, programmable quantum computers have recently emerged from decades of research investment seeking to identify new computational models which can overcome the physical limits of traditional digital technologies. Quantum computers are potentially capable of calculating solutions to problems that are simply intractable on conventional computing architectures, delivering a true quantum advantage. Since May 2016 when programmable quantum computers were first made available on the cloud [1], we have established an aggressive roadmap for scaling Quantum Volume [2], much as Moore's Law did for the performance and density of semiconductor transistor technology. Our ambition is to double Quantum Volume annually, and as a result, future computing systems will need to integrate these rapidly developing quantum computing capabilities in the upcoming years.

The cloud computing paradigm is also driving considerable disruption in computing. Cloud is already demonstrated as a principal computing delivery mechanism for the future, meeting the need to provide geographically distributed, complex compute capability across the connected global economy. Cloud also provides enterprise-level security and provisioning solutions that most enterprises cannot afford in-house. Furthermore, it supports the distributed data requirements we face ahead; data scales are growing so rapidly that compute will need to follow the data, rather than data being collected to a few aggregation points in private data centers.

Underlying all of these technology challenges and opportunities is the availability of data at an unprecedented scale and fidelity. Every element of research, enterprise, consumer, and government activity is now generating and storing data at extreme scale. This data is the foundation on which our emerging computing opportunities are based. In simple applications, AI applied to data can speed traditional analytics and modelling by orders of magnitude. Future computing architectures also hold the promise of finding solutions to complex problems directly within data, for example via Accelerated Discovery. Designing architectures to take advantage of the large, complex, and often confidential data sets needed for future applications poses multifold challenges, which include security, encryption, privacy, provenance, management and performance.

**Key technology components of future computing systems**

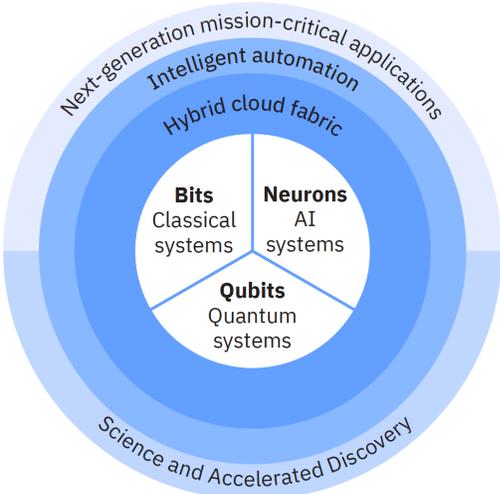

*Figure 1: The Future of Computing will weave together heterogeneous technology components to accelerate new applications.*

Our vision for the Future of Computing will weave together the advantages of emerging and complementary roadmaps to build differentiated computational capabilities. As illustrated at the center of Figure 1, the Future of Computing will be built on a foundation of bits + neurons + qubits. These heterogeneous components will be orchestrated and deployed by a Hybrid Cloud Fabric that masks complexity while enabling the secure use and sharing of both private and public systems and data. Sophisticated applications will be enabled through a layer of Intelligent Automation that combines AI and automated programming to write software code based on user needs and desired outcomes, and is optimized utilizing feedback learned from instrumentation of the computational workflows themselves. These systems will support Next-Generation Mission-Critical Applications as well as new use cases that automate complex, open-ended tasks such as Science and Accelerated Discovery. Moreover, synergies are emerging through early use cases which judiciously combine subsets of bits, neurons, and qubits in creative and powerful ways. The Future of Computing is taking shape.

## 2. Bits: Mathematics + Information

The bits used by classical computers emerged in the mid-20th century, when mathematics and information were combined in a new way to form information theory, launching both the computer industry and telecommunications. The strength of bits lies in their reliably binary nature. A bit is either zero or one, a state that can be easily measured, computed on, communicated, or stored. When provided with the same binary input, classical programs will always produce the same output. This has enabled us to build incredibly robust and reliable systems for handling high-volume enterprise workloads like transaction processing and data storage. In the Future of Computing, classical systems will serve as engines for large-scale mathematical and logical computation as well as persistent stores of data.

Moore's Law [3] and Dennard scaling [4] have provided the governing rhythm for semiconductor innovation; a doubling of transistor density, paired with performance and efficiency gains. The cadence of new technology nodes has slowed in the past decade, as the complexity and cost of innovations required to achieve density, performance, and power consumption targets has increased.

The first four decades of Moore's Law saw incremental changes in devices, materials, and process control to generate radical gains in scaling. In the last decade, and even more so for the decade ahead, radical changes in devices, materials, and process control are required for incremental progress in scaling. Technological sea changes in silicon CMOS over the past 25 years have given rise to the introduction of copper metallurgy, hafnium-based dielectrics, high-mobility strained channels, multiple patterning, Extreme Ultraviolet lithography (EUV) and the three-dimensional transistor, enabling the scaling shown in Figure 2.

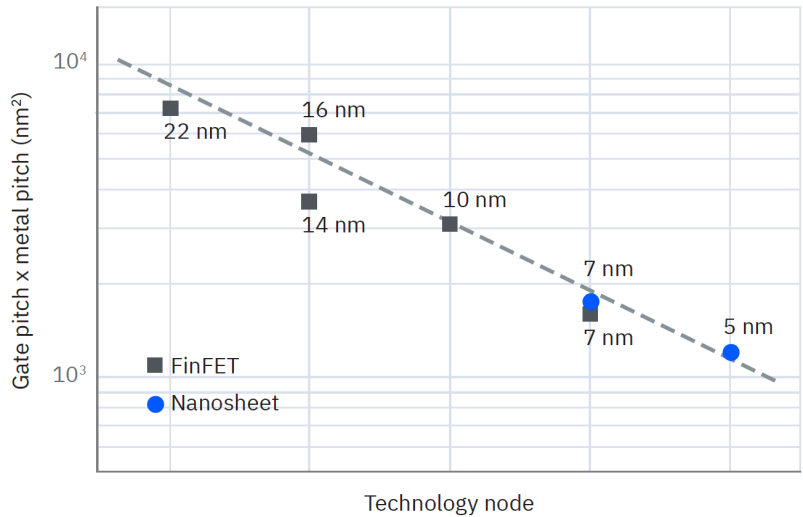

*Figure 2: Roadmap of technology node innovations for classical computing.*

The semiconductor industry saw two major crises in the past twenty years. The first crisis occurred in the early 2000s when silicon-dioxide based gate dielectric stopped scaling. Engineers came up with a material innovation introducing high-κ/metal gate to replace silicon-dioxide/poly gate. The second crisis occurred in the early 2010s when power density and dissipation became too difficult to manage in planar transistor technology. In this case, engineers overcame the challenge with a device architecture innovation introducing a three-dimensional transistor, the FinFET (Figure 3 left), to replace planar transistor. In each case, it took the industry more than ten years in R&D to overcome these challenges, whether introducing a new gate dielectric material, or a new transistor architecture.

**FinFET cross section**        **NanoSheet cross section**

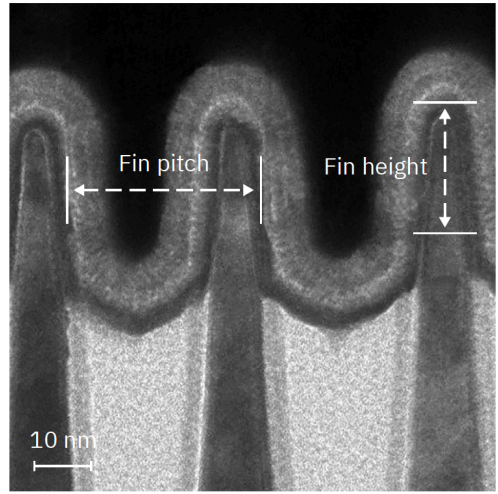 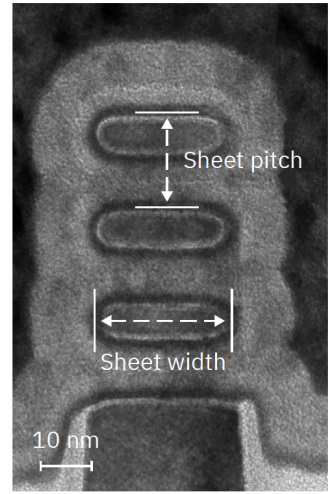

*Figure 3: Transmission electron microscope cross-sections of a FinFET device (left) and a NanoSheet device (right).*

FinFETs benefit from getting smaller (Fin width), taller (Fin height) and closer together (Fin pitch). Taller fins provide more drive strength per unit area; fins that are closer together have lower parasitic capacitance, and of course occupy a smaller area. However, there are practical limitations on how tall and how close the fins may be. Now, we see the end of FinFETs coming as the technology is approaching the barriers of Fin dimensional scaling and the associated device physics, including constraints in short channel electrostatics and carrier transport.

Gate-all-around stacked NanoSheets (NS, Figure 3 right) improve upon these aspects and can meet the needs of logic devices at a fully scaled 5 nm technology node and beyond [5]. The sheet-to-sheet spacing, analogous to Fin pitch, is determined not by lithography but tightly controlled epitaxial processes. The effective device width per area is larger than FinFETs when multiple NanoSheets are stacked. In Figure 3, an exemplary NanoSheet device is shown having three sheets to maintain a practical height. However, more or fewer sheets is in principle possible. The final sheet thickness is chosen to enable the required electrostatic behavior without encountering undue quantum and scattering effects. Compared to mainstream FinFET architectures, NS offers excellent electrostatics and short channel control, can be fabricated with limited deviation from FinFET processes, and circumvents some of the critical patterning and material challenges associated with scaled technologies. Together with variable NS width design enabled by introduction of EUV, NS device architecture can provide density, power and performance scaling benefits for CMOS logic technology beyond FinFETs.

A pipeline of architecture, materials, and patterning innovations paves the path ahead for bits. The cadence of adoption will follow the market appetite for each node. In the Future of Computing, the role of classical bit-based computing elements is evolving, but not diminishing in its criticality. The "bit player" will have a central role, conducting a computing symphony enriched by AI and Quantum systems.

## 3. Neurons: Biology + Information

Inspired by neuroscience and combining biology and information, deep learning systems have been engineered to meet and even surpass human performance on many perception-related cognitive tasks. These systems use mathematical neural networks to create AI models. These AI models are trained by learning from large data sets of examples.

Despite significant progress, we are still in the early phase of AI, known as Narrow AI, which requires vast volumes of labeled data in order to learn single, specialized tasks. While we are still far from achieving artificial general intelligence, we are now entering the era of Broad AI, which will be characterized by AI that can scale from task to task and across disciplines, and which successfully combines learning and reasoning. The creation of Trusted AI – AI that is fair, explainable, robust, and transparent – is also a central objective for enterprise and societal uses of AI [6]. These AI systems will not only learn to perform complex tasks in future applications. They will also be used to intelligently automate the design, programming, orchestration, and management of future computing systems.

Narrow AI is computationally limited in many cases; implementations of Broad AI will only exacerbate these computational challenges. Both algorithmic and hardware innovations are required to reduce the time and power consumed to create, deploy and scale increasingly complex models and networks. Creating optimized systems for these AI workloads requires re-thinking how we innovate the end-to-end system, including materials, devices, hardware, software, and programming.

The first foundational progress in compute efficiency for AI model training and inference has been made by exploiting the statistical and approximate nature of deep learning algorithms, leading to reduced-precision digital approaches [7] [8]. Floating-point multiplication dominates deep learning training, and multiplier complexity is quadratic with operand size; e.g., 16-bit precision engines are more than 4x smaller than 32-bit precision engines. This gain in area efficiency boosts performance and power efficiency. Simply stated, in approximate computing, we can trade numerical precision for computational efficiency, provided we also develop algorithmic improvements to ensure uncompromised iso-accuracy [9] [10]. We have recently demonstrated the success of this approach for training with 8-bit floating-point numbers, using new techniques to address the challenge of maintaining fidelity of the gradient computations during back-propagation [11] [12]. Likewise for deep learning inference, we illustrated the possibility of achieving state-of-the-art classification accuracy comparable to full precision networks across a range of popular models and datasets, using only 2-bit integer arithmetic [13]. We believe that precision requirements for both training and inference will continue to be driven down through a sustained stream of innovations in algorithms and hardware, leading to dramatic increase in deep learning compute capability over the next decade [14].

The next step in the evolution of specialized hardware for AI is rooted in addressing the performance efficiency loss from data movement between computational units and memory. Analog accelerators avoid the von Neumann Bottleneck by performing the computation in place. As shown in Figure 4, the analog approach uses arrays of non-volatile, programmable resistive processing units (RPUs) that encode the neural network weights. Computations like vector matrix multiplication (MAC) or element-wise matrix updates can be done in parallel and in constant time, with no movement of the weights.

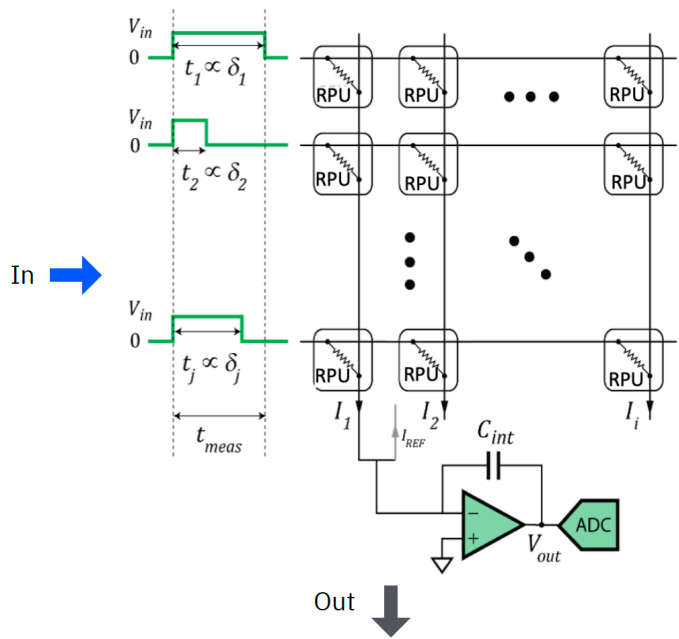

*Figure 4: A row input vector $V_{in}$ is time encoded and the column current is integrated. The accumulated charge value can either be digitized or be transferred to the next layer in the neural network as a time encoded signal.*

In an analog chip architecture, the AI network is then represented by stringing arrays together. Non-linear activation functions are inserted in the connection between arrays and can be done either in the digital space or in analog as well [15] [16]. As the weights are stationary, data traffic is significantly reduced which mitigates the von Neuman bottleneck. However, in contrast to digital solutions, analog AI alternatives will be more sensitive to material properties and are intrinsically susceptible to noise, drift, mismatch and variability [17]. These factors must be addressed through architectural, circuit, and algorithmic solutions. Mitigation of these deficiencies has progressed through incorporation of noise and bound management components, and by embedding drift compensation into the algorithms for maintaining floating-point accuracy [18] [19] .

Although a complete hardware verification of these techniques is still pending, the feasibility of analog computing for AI has been demonstrated [20]. For training, the switching behavior of the non-volatile memory (NVM) device is of crucial importance. A proof of principle is shown in Figure 5. In this experiment the weight data resides in a phase-change memory (PCM) memory hardware array and a 3-layer fully connected network is trained on the noise-augmented MNIST data set (MNIST-back-rand). To address the deficiencies of the PCM material, a modified synaptic cell separating the trailing and leading digits of the weight onto a capacitor and PCM, respectively, is introduced. The analog network test results show an accuracy that is on par with a purely digital floating-point solution using the TensorFlow framework. Another solution to mitigate the properties of the PCM material is the mixed precision approach, in which MAC is performed in the analog domain and the weight update is done in a high-precision digital environment [21]. Moreover, materials other than PCM, such as high-κ transition-metal oxides for resistive RAM (RRAM) [22] [16] are the subject of active investigation as possible NVM candidates having more ideal characteristics for analog computing.

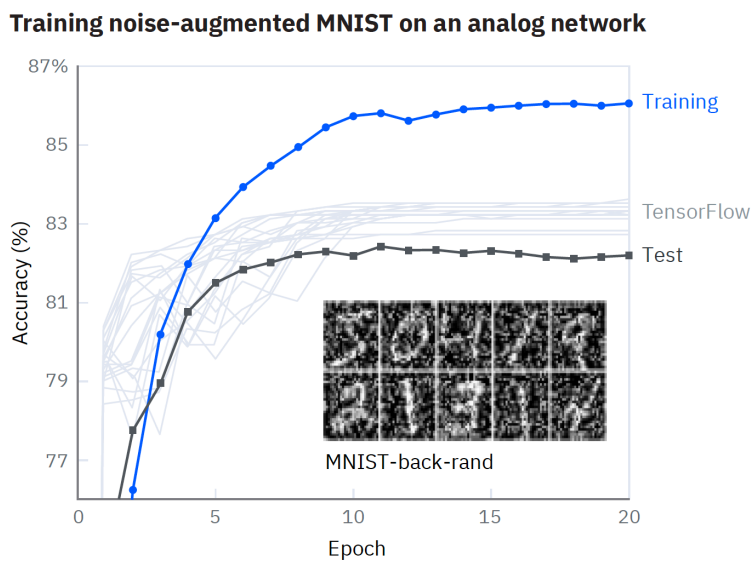

*Figure 5: Analog training and test accuracy are shown with the blue and black lines, respectively. Gray lines show test accuracy for digital floating-point computation using TensorFlow.*

Algorithmic innovations can also be employed to mitigate material imperfections and improve the performance of analog AI hardware. Figure 6 shows the results of a simulation of a 3-layer convolutional analog network training on MNIST. The gray line shows a high test error for conventional stochastic gradient descent (SGD) when a realistic switching model for a bidirectional RRAM device (shown in the inset) is incorporated into the RPU simulation. The blue line shows how a modified SGD or back propagation algorithm (referred to as the "Tiki-Taka" method) can successfully accommodate the non-symmetric RRAM switching behavior [23]. With the expense of additional training epochs, the Tiki-Taka method can reach a test error nearly identical to the fully digital floating-point baseline (gray circles). For reference, the performance of conventional SGD with a perfectly symmetric NVM device is illustrated by the black line. These examples show that analog computing can reach accuracies on par with floating-point results.

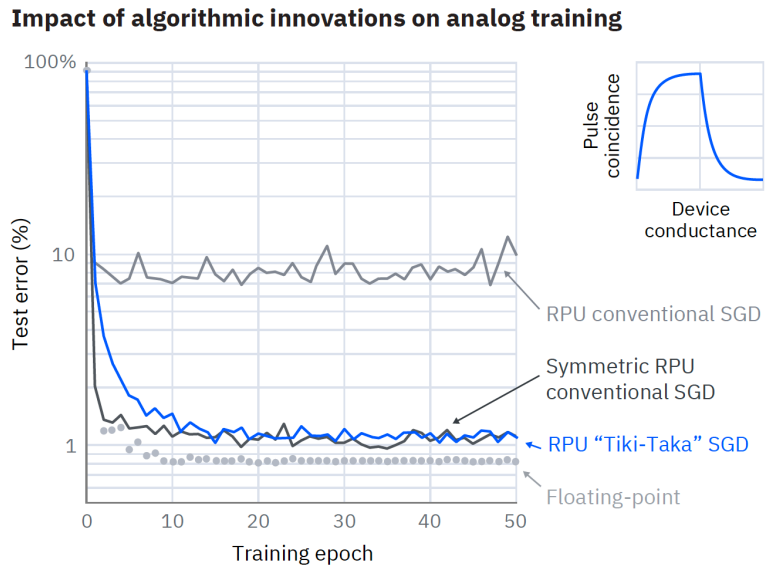

*Figure 6: Simulation of a 3-layer analog RPU network training on MNIST, comparing conventional SGD (gray line) and modified "Tiki-Taka" SGD (blue line), both with a realistic RRAM switching characteristic (inset), conventional SGD with a symmetrically switching NVM device (black line), and digital floating-point results (gray circles).*

The challenge ahead is to show that this new technology can be scaled to relevant larger networks with tens of millions of parameters and tens of layers. Initial system-level simulations show that even for inference only, building a model for analog hardware needs to take into account the above-mentioned material properties in order to obtain the weights required to maintain floating-point classification accuracy, albeit only in the forward pass [24]. Both open loop [25] and closed loop [26] weight transfer algorithms capable of transferring weights onto analog arrays without loss of accuracy have been demonstrated.

In the AI hardware roadmap shown in Figure 7, we are focused on hardware, software, and algorithm innovations that will meet the evolving and increasingly complex demands of neural network architectures. Each of the core technologies described above can be applied flexibly to training and inference, across diverse and changing workloads. We project an annual improvement of 2.5x in effective performance efficiency, and envision that these gains will be secured using the approximate computing principles applied to both digital and analog AI cores. The additional uplift shown for analog AI cores is due to the elimination of the von Neumann bottleneck for data transport, and to the computational time scaled proportionally to the network dimension, n, whereas digital computational time scales with $n^2$. Achieving the full potential of analog AI cores will require materials innovations. In this simplified roadmap, all datapoints are normalized to 16-bit precision, although advances in reduced precision raise the performance capabilities of all technologies which can achieve iso-accuracy. The methodology for the analog projections has been previously described [17], and assumes full utilization is achieved with optimized pipelining, and that high on-chip bandwidth is provisioned with heterogeneous integration to ensure an unconstrained data supply. Researchers are also developing the software toolkits required to transparently map deep learning workloads to future digital and analog AI hardware accelerators [27].

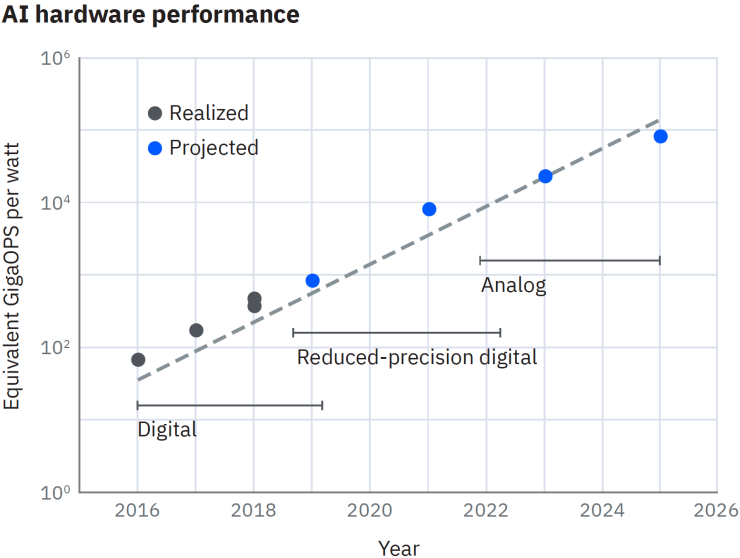

*Figure 7: Projected performance of AI hardware accelerators. Forecast to sustain 2.5x/year in equivalent GigaOPS/W is shown with blue points [17], while the gray reference points are taken from realized commercial GPU and ASIC designs [28] [29].*

## 4. Qubits: Physics + Information

Quantum bits – or qubits – combine physics with information, and are the basic units of a quantum computer. Quantum computers employ qubits in a model of computation that is based on the laws of quantum physics. Suitably designed quantum algorithms are capable of solving high-complexity problems by exploiting quantum superposition and entanglement to access an exponential state space, and then amplifying the probability of calculating the correct answer through constructive interference.

Classical computer systems built from bits scale in a linear fashion, where going from N bits to 2N bits lets you process or store twice as much information. In contrast, qubits scale with an exponential: $2^N$, where N is the number of qubits. It only takes one additional qubit, for a total of N+1 qubits, to double the computational power of a quantum computer. Within the next decade, we predict that this exponential scaling will allow quantum computers to answer technologically-relevant questions that are intractable for classical computers [30] [31] [32]. Quantum systems will be the engine accelerating the most challenging computational tasks at the heart of future applications – from solving the

Schrödinger equation controlling the inner workings of materials and catalysts, to quantum random walks for risk modeling in finance.

A proven approach to quantum computing relies on Josephson junction transmon qubits [33], assembled into quantum processor chips such as the one shown in Figure 8. The qubits are themselves constructed from and interconnected by superconducting circuits (capacitors, inductors, and microwave resonators), and support a universal gate set allowing all logical operations [34]. Quantum algorithms are run on quantum processors by applying a series of specific qubit rotations and two-qubit controlled operations. These operations are induced through application of appropriately timed and shaped microwave pulses to the cryogenically cooled physical devices.

The Josephson junction at the core of the transmon qubit consists of a superconducting electrode, a thin insulator and another superconducting electrode. Figure 8 also shows one example of a Josephson junction which has been fabricated using a crossed Dolan bridge [35]. In its superconducting state, the junction allows Cooper pairs to tunnel across the insulating barrier, and behaves effectively as a non-linear inductor. While a typical transmon qubit occupies an on-chip footprint of 600 μm x 600 μm, the Josephson junction is much smaller, with an area of only 100 nm x 100 nm. Readout of the qubit state is facilitated using a dispersive interaction with a coupled superconducting microwave resonator [36]. The resonator frequency is interrogated with a microwave pulse, and the phase and amplitude of the reflected signal are used to distinguish the state of the qubit.

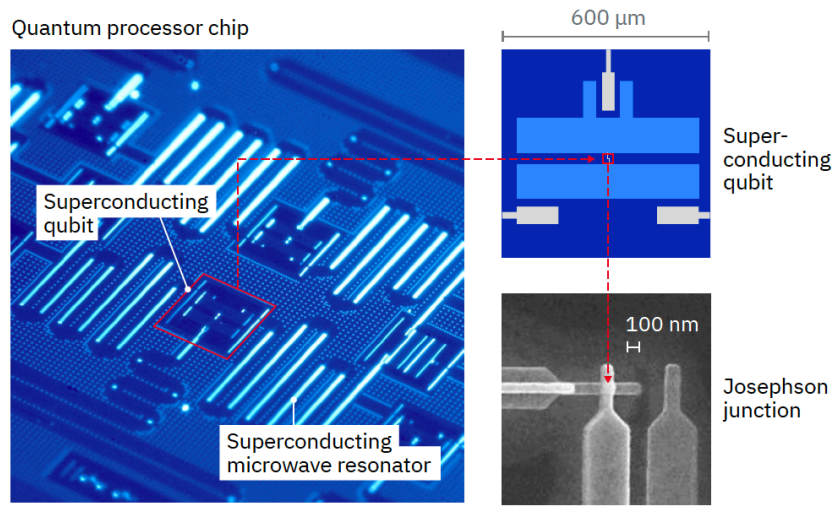

Figure 8: A quantum processor chip composed of a lattice of superconducting qubits, interconnected by superconducting microwave circuits and resonators. The Josephson junction at the center of the 600 μm x 600 μm qubit has an area of only 100 nm x 100 nm.

Important metrics for individual qubits are the energy relaxation time $T_1$, and the dephasing time $T_2$, which collectively describe the qubit coherence. Energy relaxation quantifies the time it takes for a qubit to decay from its excited state |1⟩ to the ground state |0⟩ (a bit-flip error), while dephasing correspond to the time it takes for a quantum superposition state |+⟩ = (|0⟩+|1⟩)/√2 to lose the phase relationship between |0⟩ and |1⟩ (i.e. a phase-flip error). Both quantities play an important role, as shorter coherence times reduce the accuracy of quantum operations. As the historical progression of coherence times shown in Figure 9 illustrates, the first experimental demonstration of a superconducting qubit, attributed to the group at NEC in 1999 [37], had a coherence time on the order of 1 ns. Since this seminal result, numerous variations of superconducting qubits and their associated superconducting circuits have been implemented, for example by adding loops interrupted by one or more Josephson junctions or by adding capacitors. Research has shown that charge noise, flux noise, undesirable coupling of the qubit's electric field to the microwave environment, and dielectric loss due to two-level systems (TLSs) at the microscopic level of materials play crucial roles in limiting the coherence time of superconducting qubits. Figure 9 also shows that over the last 20 years, the community's understanding of these fundamental mechanisms and the resulting innovations in qubit design [38] [39] [40] [41] [42] [43] have given rise to tremendous coherence time improvements of nearly 6 orders of magnitude. Superconducting transmon qubits having a coherence time on the order of 500 μs can be manufactured repeatably today [44].

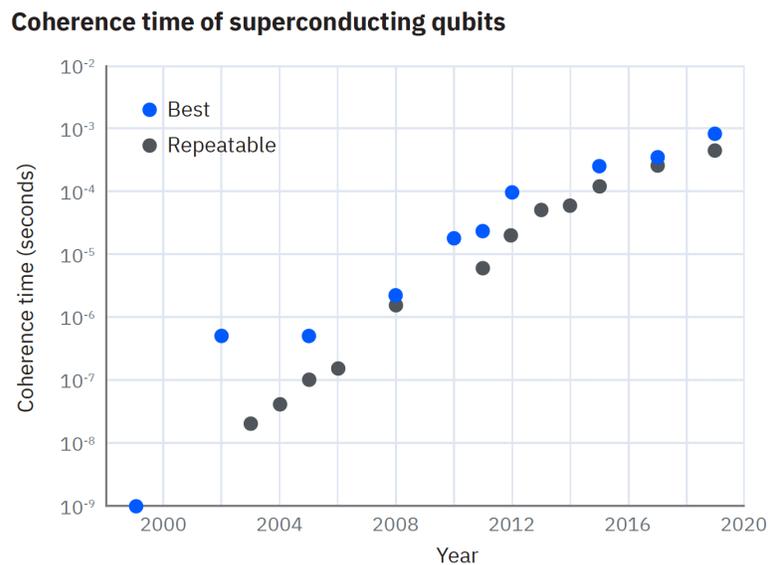

*Figure 9: Historical improvement in coherence times of single superconducting qubits.*

While improvement to the state-of-the-art in qubit coherence time remains essential, recent quantum computing efforts have moved well beyond studying simple one- and two-qubit gate characteristics, and are now focused on controlling systems with several tens of qubits [1] [45] [46]. From the first programmable five-qubit device first made available on the cloud in 2016 [1], to the 53-qubit device which was made available in October 2019 [47], IBM has taken the lead in moving the technology out of the research lab and putting it directly into the hands of developers, researchers, and quantum technology and service providers.

Along the way, the research necessary to scale the size, performance, and stability of quantum processors has led to significant improvements in the quality and uniformity of multi-qubit systems. For example, Figure 10 plots the Controlled-NOT (CNOT) two-qubit gate error distributions measured between all available connected pairs of qubits, within four successive generations of 20-qubit quantum processors [48]. Each row corresponds to a different 20-qubit device. Beginning with earlier devices (top rows), the error rates had broad distributions and a large average value. Connectivity can have a large impact upon these error rates, and accordingly on the performance of quantum algorithms executed on these devices. On one hand, more connectivity enables implementation of quantum circuits that entangle the qubits in fewer steps. However, this can come at the price of increased gate error rates or introduction of spectator errors [49] [50], i.e., errors that can occur on qubits that are still passively connected but otherwise not directly involved in a particular quantum operation. More recent connectivity map revisions (bottom rows) have sought to co-design the desired application-specific quantum circuit hand-in-hand with the quantum processor's physical connectivity, in an effort to evolve toward more optimized designs. In concert with ongoing device research to address qubit imperfections, the result has been a general reduction of error rates, tighter error rate distributions, and improved repeatability across gates.

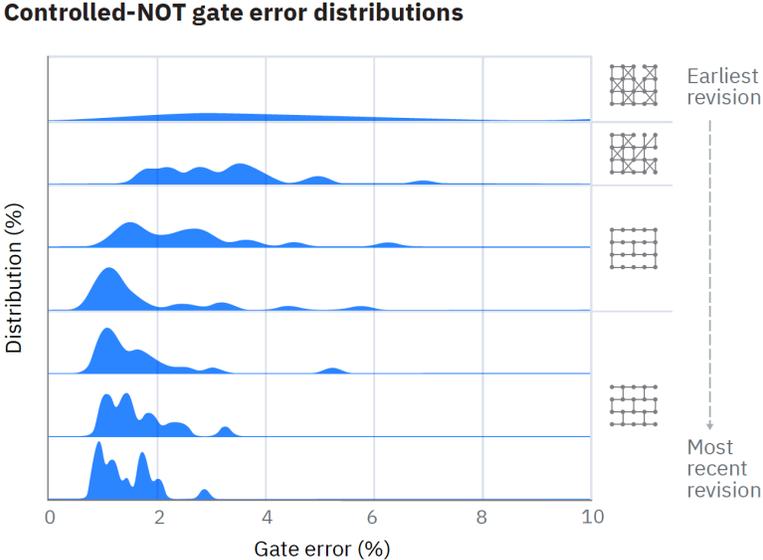

*Figure 10: Controlled-NOT gate error distributions for several generations of 20-qubit quantum processors having different connectivity maps.*

Because quantum information is fragile, quantum error correction (QEC) codes will ultimately be required to perform error-free or fault-tolerant quantum operations. As shown in Figure 11, future machines will have a physical layer that provides the error correction, and consists of a physical quantum processor that has both input and output lines that are controlled by the QEC processor. This processor is in turn controlled by the logical layer, where the logical encoded qubits are defined from many physical qubits, and in which the logical operations are performed to execute the desired quantum algorithm. However, at the present time, the overhead of constructing a fault-tolerant quantum processor is very large, and will require significant research investment to reach.

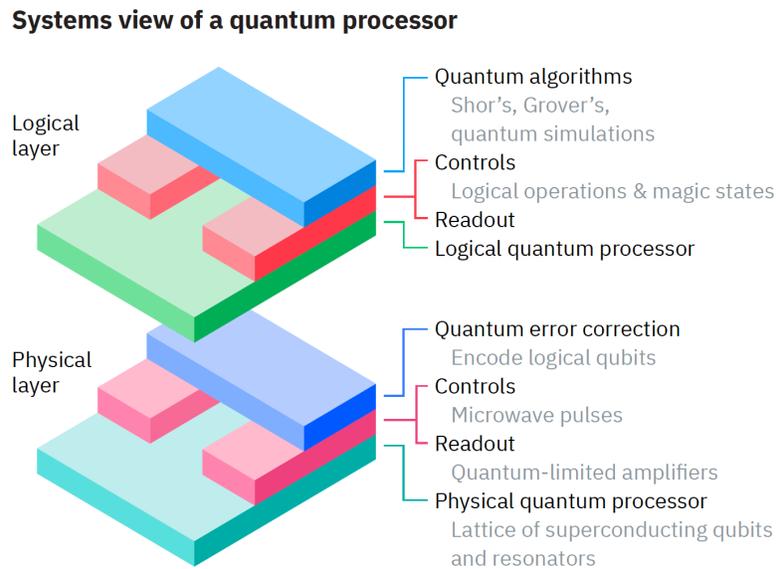

*Figure 11: A systems view of a quantum processor, consisting of a physical layer and a logical layer.*

Prior to the availability of fully error-corrected quantum computers, it is nevertheless relevant to ask if a clear quantum advantage can be gained with near-term quantum processors. A quantum advantage is obtained when for certain technologically-relevant use cases, a quantum computer can definitively demonstrate a significant performance advantage over today's classical computers. The relevance of a quantum computer is derived from the algorithms that it can execute. Not all quantum algorithms can lead to a quantum advantage. If a quantum algorithm can be efficiently simulated on classical hardware [51] [52] [53], it cannot provide a computational advantage. The advantage of a quantum computer is based on the complexity of the algorithms it can execute, and not on its ability to perform fast operations. It is therefore paramount to ensure that the quantum algorithm is based on a circuit that cannot be efficiently simulated on a classical computer.

Today, quantum circuits are practically confined to a width defined by a small number of qubits, and a shallow depth, or number of sequential computational steps, limited by qubit noise and finite coherence time, as shown in Figure 12. A fundamental question is therefore whether and under which circumstances such a shallow depth quantum circuit can provide a computational advantage. This question was recently answered through a proof of an unconditional separation in computational power between certain classes of shallow quantum and classical circuits [54]. It follows that a systematic approach to developing quantum applications which can be supported by near-term quantum devices with a reliable quantum advantage, should be based upon the complexity theoretic hardness of quantum circuits, combined with an effort to identify and map technologically relevant algorithms to these classes of short depth circuits. An example of one such algorithm, combining "qubits + neurons" to implement a binary classification using supervised training, is presented further below. A variety of additional industry use cases, including the development of advanced battery materials, molecular simulations with high chemical accuracy, product recommendation systems, maritime routing, and financial portfolio optimization, also stand to benefit from quantum computing over the upcoming decade.

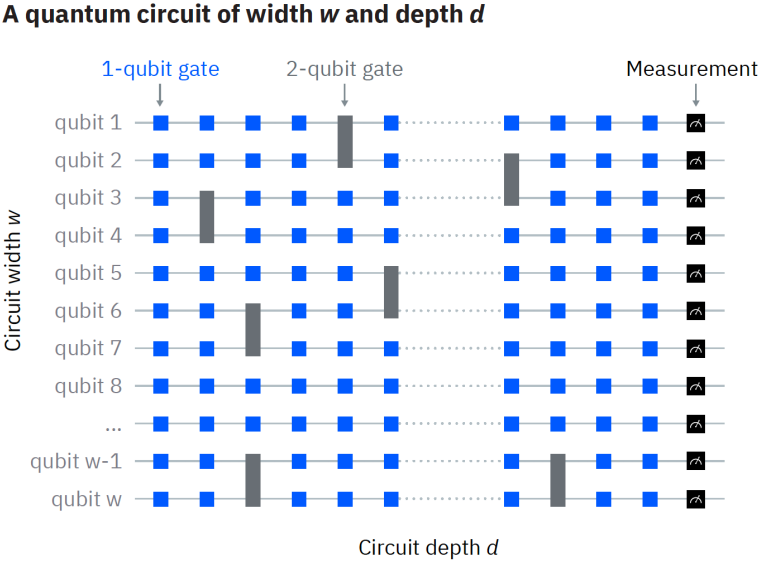

*Figure 12: Quantum circuits execute algorithms by applying a series of specific qubit rotations (1-qubit gates) and two-qubit controlled operations (2-qubit gates), followed by measurement of the qubit states. Today's quantum circuits are confined to a width given by the small number of available qubits, and a shallow depth limited by qubit noise and coherence time.*

As the technical community builds larger quantum systems capable of performing more complicated algorithms and executing deeper quantum circuits, it is important to benchmark their overall computational power. The Quantum Volume [2] is an architecture-neutral system-level performance metric that accounts for gate and measurement errors, device cross-talk and connectivity, and circuit compiler software efficiency. Improvements in Quantum Volume correlate with the ability to solve larger, more complex problems across a range of disciplines.

Figure 13 plots the Quantum Volume demonstrated across several generations of our quantum computing systems, code-named Tenerife, Tokyo, and IBM Q System One. In early 2019, IBM Q System One was benchmarked at a Quantum Volume of 16 [44]. The systems illustrated in Figure 13 have established a roadmap for Quantum Volume scaling, much as Moore's Law did for semiconductor technology. Our goal is to double Quantum Volume annually, and we anticipate that realizing this ambitious roadmap will be necessary in order to achieve quantum advantage in the 2020s.

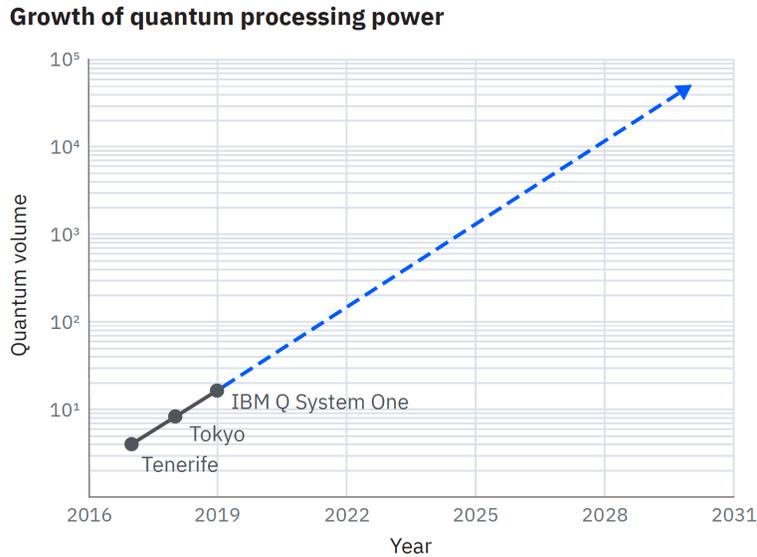

*Figure 13: Exponential forecast for growth of quantum processing power, as reflected through the Quantum Volume.*

## 5. Emerging Synergies

The Future of Computing has already begun to take shape inside our labs. The ultimate objective of designing systems and computational workflows which fully leverage the complementary capabilities of bits + neurons + qubits will remain a topic of active research for many years to come. However, several advanced computational systems and use cases are already operating at the intersections between pairs of these computational methods. The examples below illustrate several synergistic combinations that highlight the promise of fully integrated systems.

Bits + Neurons

Today we are already creating systems that combine bits and neurons. The Summit supercomputer [55], built by IBM and shown in Figure 14, is the world's most powerful "bits + neurons" system. With a peak performance of over 200 petaflops, Summit holds the #1 position on the TOP500 supercomputers list.

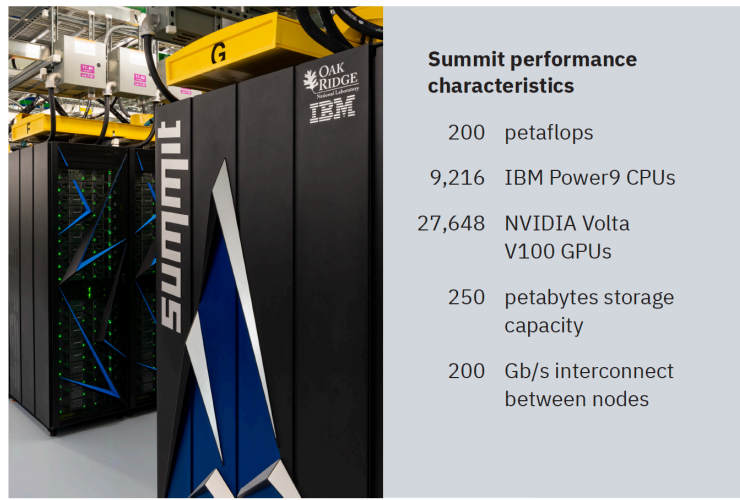

Figure 14: The Summit supercomputer at Oak Ridge National Laboratories has been ranked #1 on the TOP500 supercomputers list since June 2018. ("The Summit Supercomputer" by Carlos Jones/ORNL, used under CC BY 2.0.)

The Summit supercomputer has been purpose-built to support complex computational and AI workloads. Summit combines over 9,000 IBM POWER9 processors (bit-based computing) with more than 27,000 graphics processing unit (GPU) accelerators well-suited for neuron-based computing, along with advanced memory, storage, and interconnects to facilitate high-bandwidth movement of data between CPUs, GPUs and node-to-node across the entire system.

Each node includes 512 GB of main memory that is used to share data within and across the many different stages in the complex workloads that are now normal in AI-enabled computation. Stages in the workloads can variously include computation, learning, inferencing and steering, and it is critical for performance and throughput that data can be shared in main node memory rather than through shuffling to storage. Additionally, in the node design for Summit, both the CPU and GPU elements can read and write each other's memory consistently and coherently. This allows different stages of the workflow running on different compute elements to seamlessly access data, and avoids the need for performance-impacting code to constantly rearrange and transfer data between and within workflow stages. Summit has established some general principles for future "bits + neurons" systems including the need for high performance memory pools to share data within and across stages, and the need for shared and coherent memory access across different computing elements to maximize performance across heterogeneous computing elements.

Using Summit, recent work has demonstrated dramatic acceleration of the time required to train a state-of-the-art video activity recognition model [56]. Known as the Temporal Shift Module (TSM), this model replaces conventional 3D convolution methods for video recognition with a light-weight algorithm that requires only the reduced computational resources of 2D convolution. TSM is therefore designed to be highly hardware-efficient, while preserving model accuracy. With such a hardware-aware model design, experiments on Summit show it is possible to scale up training and reduce the training time on the Kinetics dataset [57] from 50 hours (using 1 Summit node, 6 GPUs) to 14 minutes (using 256 Summit nodes, 1,536 GPUs). The top-1 accuracy achieved by these models is 74.1% and 74.0%, respectively, demonstrating that no loss in accuracy is incurred in exchange for the 200x productivity enhancement.

Ultimately, Summit and other heterogeneous special-purpose systems like it will leverage the advantages of their "bits + neurons" design to accelerate technologically relevant workloads and deliver new scientific insights.

Qubits + Neurons
It is anticipated that quantum computing's tremendous processing power has the potential to unleash exponential advances in artificial intelligence. AI systems thrive when the machine learning algorithms used to train them are given massive amounts of data to ingest, classify and analyze. The more precisely that data can be classified according to specific characteristics, or features, the better the AI will perform. Quantum computers are expected to play a crucial role in machine learning, including the crucial aspect of accessing more computationally complex feature spaces – the fine-grain aspects of data that could lead to new insights.

We have investigated how qubits can be applied to and paired with neuron-inspired machine learning algorithms to derive a quantum advantage over classical computational approaches. Recent work [58] has demonstrated a binary classification algorithm which exploits a quantum enhanced feature space and achieves classification rates up to 100%. The principle of this work, akin to the Support Vector Machine (SVM) approach, is to implement a non-linear feature map that brings the data to classify into a space in which it can be linearly separated. A quantum processor utilizes the high-dimensionality of quantum systems in order to separate the data. A feature map believed to not scale efficiently on purely classical machine learning systems is implemented on a shallow depth quantum circuit of the class exhibiting a quantum advantage [54], described earlier. This circuit is then tested upon synthetic datasets (chosen as illustrative examples) that are separable by such a feature map within the higher-dimensional space of qubit states.

**Classification by a two-qubit quantum processor**

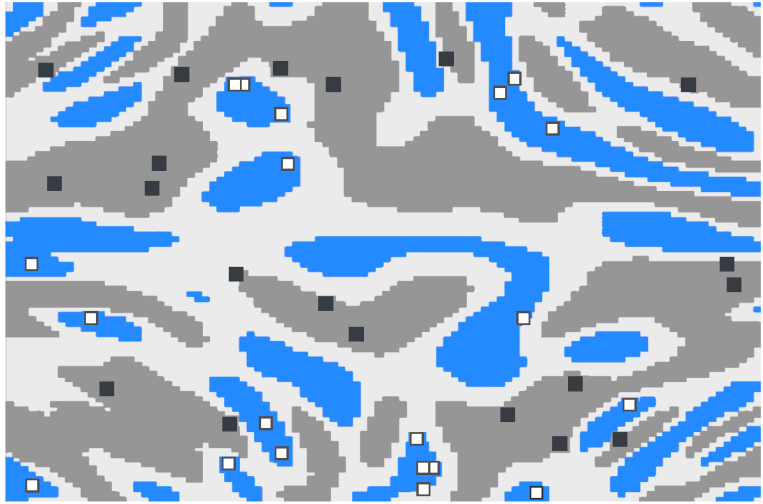

*Figure 15: A two-label (dark gray and blue) classification as performed by a quantum SVM method, where the classified test data is shown by white and black squares.*

Figure 15 shows a two-label (dark gray and blue) dataset classification as performed by a two-qubit quantum processor. The labeled data are separated by dataless light gray regions. After a training phase, the quantum processor accurately classifies individual test data points (black and white squares) as belonging to the dark gray or blue labels, respectively. Figure 16 shows the classification results of the quantum SVM approach, as a function of the two-qubit circuit depth. The gray dots represent a total of 60 classifications per depth, with mean values represented by blue dots. An increase in classification success with increasing circuit depth is observed, with the mean values reaching very close to 100% for circuits of depth larger than 1. As quantum computers become more powerful and their Quantum Volume increases, "qubits + neurons" can be combined to perform feature mapping on much more highly complex data structures, at a scale which could potentially be far beyond the reach of the most powerful classical computers.

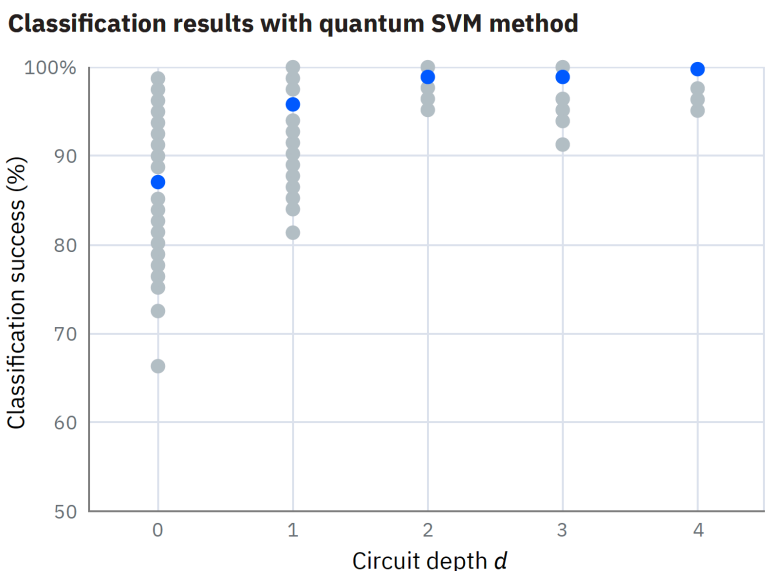

*Figure 16: Classification success improves with circuit depth, reaching close to 100% for depths larger than 1.*

Qubits + Bits

The best simulations of molecular electronic structure today rely upon classical computers which use complex methods to estimate the lowest energy molecular ground state. Knowledge of the ground state dictates the structure of the molecule and how it will interact with other molecules, and is therefore critical for chemists to design new molecules, reactions, and industrial chemical processes. However, finding exact solutions to such problems numerically has a computational cost that scales exponentially with the size of the molecular system. Therefore, even high-performance computers are eventually restricted to using approximate methods which limit accuracy. By contrast, the exponential state space accessible to fully error-corrected quantum processors has the potential to overcome these scaling limits and unlock the ability to perform fully accurate electronic structure calculations.

In near-term quantum computers, we have developed a variational quantum eigensolver (VQE) [59] [60] employing short depth quantum circuits, which combines "qubits + bits" to build a hybrid quantum-classical algorithm [61]. In this approach, molecular energy computations that are known to scale unfavorably on classical hardware are performed using qubits, while optimization routines (such as gradient descent, for example) that supply new sets of molecular "trial state" parameters are run on classical bits-based computers. Figure 17 illustrates the VQE algorithmic sequence. First, a molecule's fermionic Hamiltonian is transformed into a qubit Hamiltonian, with an efficient mapping that reduces the number of qubits required in the simulation. A hardware-efficient quantum circuit that utilizes the naturally available gate operations in the quantum processor is used to prepare trial ground states of the Hamiltonian. These trial states depend upon a set of classical parameters $\theta$. The quantum processor is driven to the trial ground state, and measurements are performed that allow us to evaluate the energy of the prepared trial state. The measured energy values are fed to an optimization routine on a classical system, which computes a cost function and generates a new set of parameters $\theta$ which will further minimize it. Iterations are performed until the parameters $\theta^*$ of the lowest energy are obtained to within a desired convergence condition.

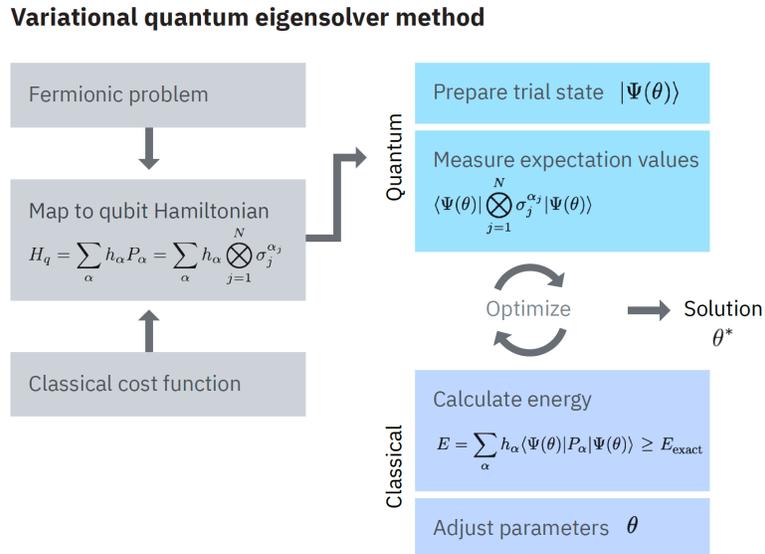

*Figure 17: Schematic of the hybrid variational quantum eigensolver algorithm, combining both quantum and classical computational techniques.*

Figure 18 depicts a simulation of the ground state energy of a LiH molecule as a function of interatomic distance, using the hybrid quantum-classical algorithm on a four-qubit processor [61]. Figure 19 depicts more recent error-mitigated results, where a method of repeating the computation at varying levels of noise is applied to generate a zero-noise extrapolation with significant improvements in accuracy [62]. This error-mitigation method is applicable to any existing quantum computer without additional hardware modifications.

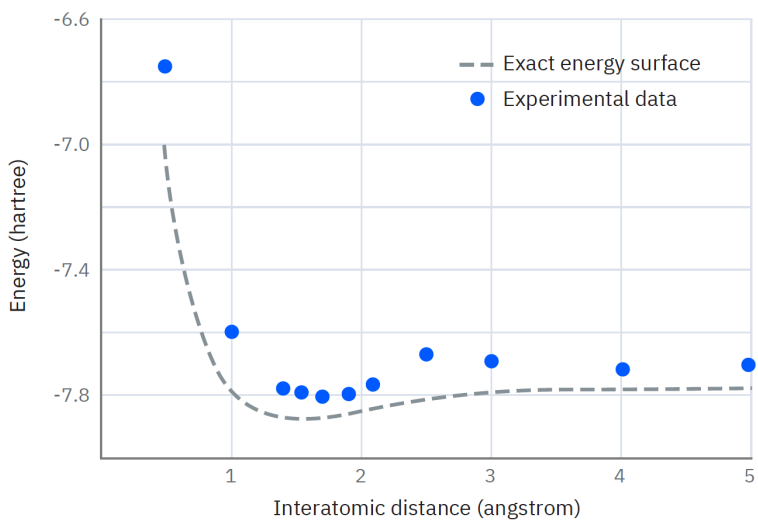

*Figure 18: Molecular simulations of LiH ground state energy using noisy quantum processors. Results from [61] without error mitigation.*

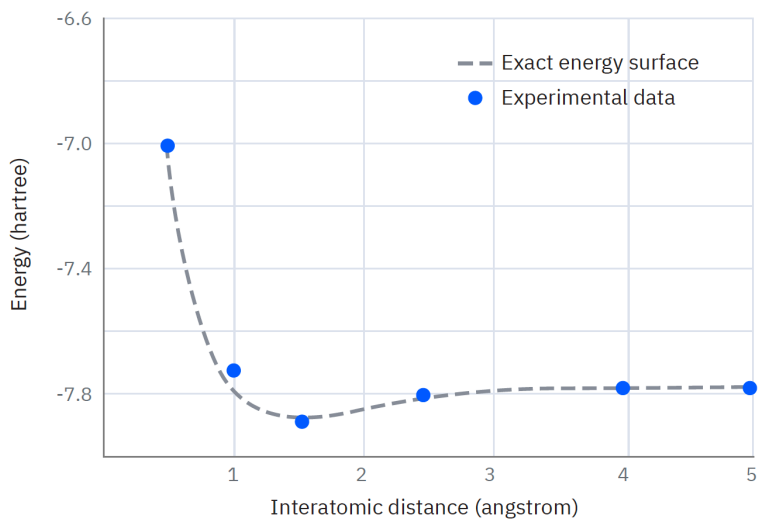

*Figure 19: Molecular simulations of LiH ground state energy using noisy quantum processors. Results from [62] with error mitigation, showing a 10x improvement in accuracy.*

As future quantum processors with larger Quantum Volume are developed, we will be able to explore the power of these and other "qubits + bits" approaches to model increasingly complex molecules, well beyond the reach of classical computing capabilities alone. The ability to simulate chemical reactions accurately will be conductive to the efforts of discovering new drugs, fertilizers, even new sustainable energy sources.

## 6. Integration of Bits + Neurons + Qubits

The paradigm of computing systems illustrated in Figure 1 will drive significant value in two highly important classes of use: Next-Generation Mission-Critical Applications, and Science and Accelerated Discovery. In comparison with today's computer systems, which are largely programmed to achieve a specific narrow task, these future applications will function more like collaborative assistants and advisors, developed to reason cooperatively with humans in ways that are natural. They will also be capable of presenting insights extracted from incredible volumes of data, amplifying human capabilities and allowing us to more effectively apply our intuition and intelligence.

As one example, we consider the field of Accelerated Materials Discovery. This field describes an overarching framework based on HPC, AI, and cloud technology, which is designed to speed up the pace of scientific discovery and technical R&D with significant reduction of time-to-market. Accelerated Materials Discovery begins where a primary human limitation lies: the large volume of data produced in all technical fields is becoming intractable by scientists and domain experts. In 2018, in the field of material science alone, there were a staggering 450,000 new publications. There is great promise in applying AI to alleviate these barriers. For example, our research has demonstrated it is possible to accurately ingest more than 100,000 PDF pages per day on a single server node, and then train AI models that extract content from these documents to create a knowledge base usable by humans [63]. Furthermore, effective knowledge representation technologies that allow experts to rapidly extract and reliably structure the information are also needed. Knowledge graph generation, simplification, and comparison technologies, as well as powerful graph inference methods, allow unprecedented fidelity in knowledge representation [64], and simplify how experts can consume the large volumes extracted. Moreover, machine learning-based surrogate models for physical systems [65] achieve more than 90% accuracy in solving complex chemical problems, assisting expert chemists in designing complex synthesis pathways. The combination of these technologies, based on "bits + neurons," has led to the development of new tools that permit domain experts to focus more upon outcomes, with the ultimate goal of speeding up groundbreaking new science and reducing the time-to-market for innovative materials.

**Cycle of Accelerated Discovery**

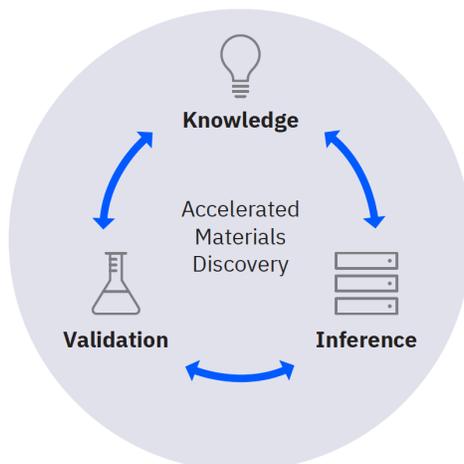

*Figure 20: The interactive cycle of Accelerated Materials Discovery, incorporating knowledge ingestion and structuring, inference and materials hypothesis generation, and numerical validation of candidate materials.*

The next integration within the Accelerated Materials Discovery framework will incorporate qubits in addition to bits and neurons. The quantum processors available today are already being applied to accurately model simple materials, augmenting traditional chemistry calculations [66]. Quantum computers can also be applied to AI tasks [58]. We envision an intelligent automated virtual experiment selector combining "bits + neurons + qubits," which ingests knowledge, identifies knowledge gaps, infers potential material candidates, and configures an optimal investigative multi-physics simulation workflow. As shown in Figure 20, such systems will be capable of interactively and iteratively evaluating and validating new materials which meet a set of desired physical traits. Within the future computing framework, these complementary technology roadmaps will enable new value, precision, and productivity within the fields of chemistry and materials.

## 7. Conclusion

Our simple, yet radical, belief is that we can weave bits, neurons, and qubits together to form a powerful foundation for the Future of Computing. This new generation of computing has already begun to take shape inside our labs. In the future, researchers and clients will use this new computing resource to quickly build and deploy creative, intelligent applications that solve complex scientific, business, and societal problems, some of which we have yet to even recognize.

We have already seen how digital and AI combine to transform the cognitive enterprise. Imagine what will be possible as we look ahead to the next decade, when we will have this new Future of Computing stack to build on. Combining bits, neurons, and qubits – and orchestrating their power using a secure hybrid cloud fabric and a layer of intelligent automation – will allow us to accelerate the rate of discovery and solve problems that truly matter to businesses and the world.

**Acknowledgements:**

This paper surveys a large cross-section of work across the IBM Research division. The authors gratefully acknowledge the contributions of Cindy Goldberg, Jeffrey Burns, Wilfried Haensch, Huiming Bu, James Sexton, Michael Rosenfield, Donna Dillenberger, Geoffrey Burr, Kailash Gopalkrishnan, Winfried Wilcke, Jay Gambetta, Markus Brink, Jerry Chow, Teodoro Laino, Costas Bekas, Alessandro Curioni, Jeff Welser, Steven Tomasco, Kathryn Guarini, Bruno Flach, Pavithra Harsha, Prasenjit Dey, Jed Pitera, John Cohn, all from IBM Research, and Professor Song Han of the Massachusetts Institute of Technology. The authors also thank Bonnie Scranton for design of the figures.